\documentclass[10pt,a4paper,twocolumn]{article}
\usepackage{authblk}
\usepackage{graphicx}

\begin{document}
\date{}

\title{Automatic Generation of Challenging Road Networks for ALKS Testing based  on Bezier Curves and Search }

\author{Florian Kl{\"u}ck}
\author{Lorenz Klampfl}
\author{Franz Wotawa}
\affil{Christian Doppler Laboratory for Quality Assurance \\Methodologies for Autonomous Cyber-Physical Systems\\ Institute for Software Technology, Graz University of Technology \\ Inffeldgasse 16b/2, A-8010 Graz, Austria}

\affil{\textit {\{fklueck, lklampfl, wotawa\}@ist.tugraz.at}}

\maketitle

\begin{abstract}
In this paper, we outline an approach for automatic generation of challenging road networks for virtual testing of an automated lane keep system. Based on a set of control points, we construct a parametric curve that represents a road network, which defines the dynamic driving task an automated lane keep system  equipped vehicle has to perform. Changing control points has global influence on the resulting road geometry. Our approach uses search to find a set of control points that results in a challenging road geometry, eventually forcing the vehicle to leave the intended path. We evaluated our approach in three different search-configurations regarding test efficiency and practical applicability for automatic virtual testing of an automated lane keep system. 
\end{abstract}

\section{Introduction}

The recent strive for developing Advanced Driver Assistant Systems (ADAS), promises improved road safety and driving comfort but also introduces new challenges to the well-established verification and validation methods in the automotive industry as described in \cite{koopman2016} and \cite{Wotawa2017}. In addition, as for every development, it is crucial to detect failures at an early stage, which requires efficient virtual testing strategies that take low preparation effort. In this paper, we contribute to this challenge, and present a method allowing to generate roads comprising curves that may challenge an automated lane keep system (ALKS). The underlying idea is to make use of a genetic algorithm allowing to searching for control points of a road that increases the likelihood of an ALKS to leave the road. In the following, we describe the method and present initial results obtained.

\section{Method Overview}

The proposed method aims to generate challenging road networks for testing an ALKS system. In ADAS testing the potential input space is near infinite. For this method we reduce the input parameter space and only consider various road geometries for testing. Road geometry only represents a subset of the static driving environment but is strongly connected to one of the most basic ALKS functionalities: controlling the lateral and longitudinal driving tasks in response to changing road conditions. In our method, a road network is represented as a parametric curve, which is constructed based on a set of control points. The road points described by the parametric curve form the input for automatic generation of virtual roads that are executable in a simulation environment. 

The resulting road network defines the dynamic driving task an ALKS equipped vehicle has to perform. Changing the location of one control point has global influence on the resulting road geometry. During testing, we search for  control point arrangements that result in challenging road networks. We consider road networks as challenging when the ALKS equipped vehicle does not perform the dynamic driving task correctly and leaves its intended road path. For search, as in our previous work \cite{klueck2019}, we use a genetic algorithm, where one set of control points represents one individual in the seed population. For evaluation of one control point individual, we construct the corresponding Bezier curve and hand over the obtained road points to an underlying code-pipeline \cite{toolcomp2020}, where the final road is constructed, validated and executed in a virtual environment namely BeamNG.research \cite{beamng_research}. 

To determine each individual's fitness, we monitor the percentage of how much the ego vehicle crosses the center line or leaves the road. During mutation, with a certain probability, we assign a new control point position within a permissible mutation range. 
We carried out an initial ALKS case study, where we evaluated the performance of our approach in three different search-configurations regarding test efficiency and practical applicability for automatic virtual testing of an ALKS. 

We are furthermore interested in finding a configuration for this approach that promotes efficient searching for challenging parameter combinations in different areas of the search space. Representing candidate solutions as a set of control-points (i.e.: a set of coordinates in a finite map of a given size) provides the possibility to keep track of already evaluated solutions. Also solutions that are very similar to already evaluated solutions should be avoided. The similarity between candidate solutions can be determined by calculating the discrete Fr\'echet Distance \cite{discrete_frechet_distance} between the two resulting Bezier curves .  The average similarity between candidate solutions in a population based on the Fr\'echet Distance can be used as indicator for the generation of new candidate solutions, for instance to only accept new solutions that increase the overall population's average Fr\'echet Distance.

\section{ALKS Case Study} \label{ALKS Case Study}

The test input provided by the proposed method is automatically processed by an underlying code-pipeline, until a certain time-budget is consumed. One test input is a sequence of points that provides the basis for the generation of an executable virtual road. The virtual road defines the driving task, which the vehicle, equipped with an omniscient trajectory planner, has to perform. We consider a test as failed, when the side or center line is crossed by more the 95 percent of the vehicle's bounding box. Otherwise a test is considered as passed. Before execution, a validity check is carried out on each test input, to assure that the resulting road does not overlap or contain any curves, which are too sharp for the vehicle to pass. In this case we consider the test as invalid. We compared the performance of our approach in three different search-configurations: 

\begin{itemize}

\item GA-Bezier Configuration A: 
In this configuration we randomly generate a seed population and continue the search until the time budget of 5000 seconds is consumed. We carried out four test runs with seven control points and a seed population containing 25 individuals

\item GA-Bezier Configuration B: 
In this configuration, once a failing test case is found, we randomly generate a new seed population until the time budget of 5000 seconds is consumed. We carried out four test runs with seven control points and a seed population containing 25 individuals.

\item GA-Bezier Configuration C: 
In this configuration, once a failing test case is found, we guide the generation of a new seed population by performing a validity check that prefers valid candidate solutions, but does not strictly exclude invalid candidate solutions from the new seed population. Since the validity check is computational demanding, we increased the time budget to 10000 seconds and decreased the population size to 15 individuals. 
\end{itemize}

\section{Results} \label{Results}

In this chapter we present the results from the performance comparison of our approach in three different search-configurations. Table \ref{tab:configuration A} shows the results obtained from four test runs carried out in configuration A.

\begin{table} [hbt!]
\caption{Test results obtained from four test runs of configuration A. T: test cases generated, P: test cases passed, I: invalid test cases, F: failing test cases.}

\begin{center}
    \begin{tabular}{ | l | l | l | l | l |p{1.1cm} |p{1.1cm} |}
    \hline
    \multicolumn{7}{|c|}{GA-Bezier  - Configuration A} \\
    \hline
    Test  & T & P & I & F & Avg. & Max.\\
    Run &    & &  &  & Fr\'echet & Fr\'echet \\
    &  &  &  &  & Distance & Distance \\  \hline
    1 & 374 & 317 & 44  & 13 & 5,23 & 6,15\\ \hline
    2 & 336 & 262 & 27  & 47 & 8,22 & 110\\ \hline
    3 & 350 & 280 & 41  & 29 & 4,82 & 37,94\\ \hline
    4 & 348 & 294 & 24  & 30 & 2,236 & 4,44\\ \hline

    \hline
    \end{tabular}
    \label{tab:configuration A}
\end{center}
\end{table}

For this configuration we found a rather high number of failing test cases in every test run. However, the average Fr\'echet  Distance between failing test cases is small, indicating that the road geometry described by each failing test case in a run is very similar. In run two and run four, we see that the maximum Fr\'echet  Distance is significantly larger compared to the average value, indicating that we found at least two distinct test cases in these runs.  Testing similar failing test cases multiple times, reduces test efficiency since we are stuck in a region of the input space that has already been covered. Therefore we also reduce the likely-hood of triggering other possible failures in the system within the remaining time budget for testing. In Configuration B we address this problem, by restarting the search every time a failing test case is found and replace the old population with a randomly generated new seed population. The results are described in Table \ref{tab:configuration B} 

\begin{table} [hbt!]
\caption{Test results obtained from four test runs of configuration B. T: test cases generated, P: test cases passed, I: invalid test cases, F: failing test cases.}

\begin{center}
    \begin{tabular}{ | l | l | l | l | l |p{1.1cm} |p{1.1cm} |}
    \hline
    \multicolumn{7}{|c|}{GA-Bezier  - Configuration B} \\
    \hline
    Test  & T & P & I & F & Avg. & Max.\\
    Run &   & &  &  & Fr\'echet & Fr\'echet \\
    & &  &  &  & Distance & Distance \\  \hline
    1 & 324 & 283 & 39  & 2 & 120 & 120\\ \hline
    2 & 321 & 269 & 50  & 2 & 88 & 88\\ \hline
    3 & 369 & 311 & 57  & 1 & n/a & n/a\\ \hline
    4 & 405 & 321 & 81  & 3 & 84 & 109\\ \hline

    \hline
    \end{tabular}
    \label{tab:configuration B}
\end{center}
\end{table}

On the one hand we found in average two unique failing test cases per test run in this configuration, which could be interpreted as a slight improvement in test efficiency. On the other hand, restarting the genetic algorithm means we have to evaluate a whole new seed population before the actual search can start again, which consumes a large part of the time budget. Further more, we see a rather high number of invalid test  cases in every test run. The contribution of these individuals in finding a failing test case is probably very low, since they are unlikely to be selected for crossing and mutation when generating a new population for the next generation. In Configuration C we address this problem, by guiding the seed generation to preferably include individuals that result in valid road geometry.  However, we do not strictly exclude individuals that result in invalid road geometry.

The results for Configuration C are described in Table \ref{tab:configuration C} 

\begin{table}[hbt!]
\caption{Test results obtained from two test runs of configuration C. T: test cases generated, P: test cases passed, I: invalid test cases, F: failing test cases.}

\begin{center}
    \begin{tabular}{ | l | l | l | l | l |p{1.1cm} |p{1.1cm} |}
    \hline
    \multicolumn{7}{|c|}{GA-Bezier  - Configuration C} \\
    \hline
    Test  & T & P & I & F & Avg. & Max.\\
    Run &   & &  &  & Fr\'echet & Fr\'echet \\
    & &  &  &  & Distance & Distance \\  \hline
    1 & 179 & 158 & 17  & 4 & 73,79 & 107\\ \hline
    2 & 255 & 220 & 29  & 6 & 58,7 & 64\\ \hline

    \hline
    \end{tabular}
    \label{tab:configuration C}
\end{center}
\end{table}

For configuration C, we significantly reduced the number of invalid test cases and also increased the number of distinct failing test cases. However the additional validity check is time consuming. We increased our time budget to 10,000 seconds and reduced the population size to 15. Therefore, the results are only conditionally comparable to the results obtained for configuration A and B.

\section{Conclusion} \label{Conclusion}

In this paper, we discuss an approach for computing the shape of roads with the purpose of challenging an automated lane keep system system. The approach makes use of Bezier curves generated using a genetic algorithm implementing search-based testing. The algorithm utilizes the car's performance for a particular generated road on crossing the center line or leaving it for evaluating the fitness. In addition, we presented first experimental results utilizing different configurations of the genetic algorithm. For all configuration and all trials the approach was able to come up with at least one failing test case.

\section*{Acknowledgment}

The financial support by the Austrian Federal Ministry for Digital and Economic Affairs and the National Foundation for Research, Technology and Development is gratefully acknowledged.

\end{document}